\documentstyle[twocolumn,epsfig,psfig]{article}
\def\e{\varepsilon}
\def\be{\begin{equation}}
\def\ee{\end{equation}}

\begin{document}

\title{Universality of the critical conductance distribution in various dimensions}

\author{I.Trav\v enec$^*$
 and P.Marko\v s$^\dag$\\
Institute of Physics, Slovak Academy of Sciences, D\'ubravsk\'a cesta 9,
842 28 Bratislava, Slovakia}

\maketitle

\begin{abstract} 

We study numerically the metal - insulator transition in the Anderson model on
 various lattices with   dimension $2 < d \le 4$ (bifractals
and Euclidian lattices). The critical exponent $\nu$ and the critical
conductance distribution are calculated. We confirm that $\nu$ depends
only on the {\it spectral} dimension. The other parameters - critical disorder,
critical conductance distribution and conductance cummulants - depend also on lattice topology.
Thus only qualitative  comparison with theoretical formulae
for dimension dependence of the cummulants is possible.  
\end{abstract}

\bigskip

\noindent PACS numbers: 71.30.+h, 71.23.An, 72.15.Rn

\bigskip
It is commonly accepted, though not proved, that metal - insulator transitions
(MIT) can be described by one-parameter scaling theory \cite{AALR}. The critical
exponent $\nu$ which describes the divergence of correlation length at MIT
depends only on the system dimension for a chosen universality class.
Microscopic details of models do not affect it. This was confirmed by
numerical analysis of quasi-one-dimensional (Q1D) systems \cite{MacKK}.
Theoretical dependence of $\nu$ on dimension $d=2+\varepsilon$ was found in \cite{wegner,wegnera}
for small $\varepsilon$. Numerically, $\nu(\varepsilon)$ was studied   on bifractals 
\cite{grussbach}.

The conductance $g$ was originally chosen as the order parameter in the
scaling theory \cite{AALR}. Soon it became clear, that the absence of self-averaging
of $g$ in the critical region must be taken into account
\cite{anderson,shapiro}. 
The shape of
the critical conductance distribution $P(g)$ in 3D models was numerically
analysed in detail 
\cite{MK,EPL,PRL,distribution,SOK}. 
Contrary to the critical exponent, $P(g)$ is not
universal. Its shape depends not merely on the dimension \cite{PRL} and physical symmetry
\cite{distribution} but also  on boundary conditions \cite{SOK} and even anisotropy 
\cite{withspiros}.
Nevertheless, for a given physical model the mean conductance and resistance
follow one parameter scaling
\cite{SMO}.

Analytical theory of MIT is restricted  to systems with dimension close to the lower critical dimension: $2+\e$ with $\e\ll 1$
\cite{theorya}. 
In spite of predicted non-universality of higher order conductance cummulants
 $\langle \delta g^n\rangle$, 
\be\label{kumulanty}
\langle \delta g^n\rangle =\left\{\begin{array}{ll} \e^{n-2} & n<n_0=\e^{-1}\\

                                                   \sim L^{\e n^2-n} & n>\e^{-1}
					\end{array}\right.
\ee
the distribution $P(g)$ should be universal in the infinite system size limit
\cite{SC}. For small $\e$, the bulk of the distribution is approximately Gaussian near
 the mean value $\langle g\rangle$. The parameters of Gaussian peak,
\be\label{edon}
\langle g\rangle\sim \e^{-1}\quad\quad
{\rm and}\quad\quad 
{\rm var} g =\langle g^2\rangle-\langle g\rangle^2\sim \e^0.
\ee
agree with the estimation of the first cummulants (\ref{kumulanty}).
For large $g$, theory predicts long power-law tail  $P(g)\sim g^{-1-2/\e}$
\cite{SC}. Numerical results for  3D  systems show 
completely different $P(g)$,   indicating that theory is not applicable
to large $\e$ \cite{MK}.

\medskip

In this Letter we  present   the critical
exponent and  the critical conductance distribution for  Anderson model 
obtained numerically on three  bifractal lattices with dimension $2<d<3$. 
They all  possess the same fractal
dimension $d_f=\log 3/\log 2+1$.
Two of these  lattices  have the same  spectral 
dimension $d_s$.  Their critical exponents are identical within error bars.
This confirms the universality of MIT.
The shape of the $P(g)$, and the value  of the critical 
disorder, depend not only on $d_s$ but  also on the lattice topology. 
This novel  non-universality of $P(g)$ is found also 
by numerical analysis of two different 3D lattices.

Topology dependence of the $P(g)$ and of the critical disorder 
disables the verification of  theoretical formulae for conductance cummulants.
Known theoretical formulae may  be valid only for $d$-dimensional
hyper-cubes which can be numerically simulated  only for integer $d$. 
Thus we calculated the
critical parameters for four dimensional (4D) lattice  and compared them with known results
for 3D lattice.  Ratios of the first two conductance cummulants are
$\langle g\rangle_{\rm 3D}/\langle g\rangle_{\rm 4D}\approx 2$
and 
var $g_{\rm 3D}\approx$ var $g_{\rm 4D}$,  in 
full  agreement with relations (\ref{edon}).

\medskip

We consider the  Anderson Hamiltonian
\be\label{ham}
{\cal H}=\sum_n\varepsilon_n|n\rangle\langle n|
+\sum_{[nn']}|n\rangle\langle n'| +|n'\rangle\langle n|.
\ee
Random energies $\varepsilon_n$ are uniformly distributed 
from  $\langle -W/2,W/2\rangle$. Parameter $W$
measures the disorder strength. 
Fermi energy equals to zero.
$n$ numbers lattice sites, and $[nn']$ are two nearest-neighbor sites.

All systems  under  consideration are linear in the $z$-direction. 
In the   plane perpendicular to the current direction we construct 
fractals A-C according to  figure 1.
A combination of a $d$-dimensional fractal with linear chain in $z$ direction produces
a bifractal  with  dimension $d+1$ both for $d=d_s$ and $d=d_f$.
The length of the system in $z$ direction is $L=2^n$ and the number of  lattice sites 
in the slice  grows as $N=3^n$ for the $n$th generation
of bifractal.  We study also  two   3D models with triangular and 
honeycomb  2D lattice in the $xy$ plane  (referred as 3t and 3h) and compare 
the obtained  results with known data for 3D cubes (3s) \cite{PRL,SOK,SMO}.

In \cite{grussbach} it was supposed that 
critical parameters are  completely determined by
the spectral dimension $d_s$ of the lattice
\cite{spectral}.   
Following \cite{spectral} and \cite{kva}
we find  the analytical values for the  spectral dimension for fractals A-C
\cite{exact}:
$d_s^A=d_s^B=2\log 3/\log 5$  and $d_s^C=2\log 3/\log 6$.
The last value differs slightly from the one  obtained by 
numerical simulation  of a random walker 
\cite{grussbach}.

Linear form of all systems in $z$ direction 
enables  us to apply the standard
numerical procedure for calculation of the critical disorder 
$W_c$ and critical exponents $\nu$ in Q1D systems
\cite{MacKK}
and  the Landauer formula for conductance  (in units $e^2/\hbar$)
\cite{ando}
\be\label{land}
g=2 {\rm Tr}~t^\dag t.
\ee

The same spectral dimension of fractals A and B requires $\nu^A=\nu^B$.
Although recently reported data, $\nu^A=2.2\pm0.2$ \cite{spiros}  
and $\nu^B\approx 2.5\pm 0.25$\cite{grussbach}, could be regarded 
as equal to each other within error bars, we wanted to check their equivalence
more  accurately.
In  our analysis  we consider first five  generations
($L\le 32$) of fractals. 
 The smallest Lyapunov exponent 
$z_1$ was calculated with accuracy of 0.1\% for L=4,8 and 16
and 0.5\% for $L=32$.
Our result indeed confirms $\nu^A=\nu^B$ (Table 1).

We tested the universality of $\nu$ also for 3D lattices with various topology. 
Q1D systems
up to  $14^2$ lattice sites in the $xy$ plane were considered. The accuracy
of the first Lyapunov exponent was 0.1\% for small crossections, and decreases as  $L$
increases, being only 
0.5-1\% for $L=14$. Our results confirm the universality of $\nu$ as expected.

For 4D hyper-cubes  we involve  systems up to  $7^3\times\infty$. Our resulting
$\nu\approx 1.1$ is in a very good agreement with the previous ones
\cite{grussbach,isa}.
The obtained critical parameters are presented in Table 2.

\medskip

The critical conductance distribution $P(g)$ for lattices A, B and C is presented in figure 2.
For bifractals A and B we can approve the system-size
independence of $P(g)$. For  C, 
we do not reach  the limiting form of  $P(g)$ because 
of finite-size effects which are much stronger at lower $d_s$.

As supposed, the mean conductance 
$\langle g\rangle$ increases as $d_s$ decreases.
We cannot, however, compare our data with the theory, 
since the theoretical analysis
has been performed only for $d$-dimensional hyper-cubes. 
We can only describe some general
features of the critical distribution:
(i) $P(g)$ converges to Gaussian as $d_s$ decreases to 2, as predicted in
\cite{SC}. 
(ii) We found no evidence of the power law decrease of $P(g)$ 
for $g\gg 1$. This could be caused
by the small statistical ensembles (we have
 only $\sim 1.000$ samples in the 6th generation, 
$L=64$).  We  note that the qualitative arguments 
against the power-law decrease
\cite{MK} may not be valid in the limit 
$d\to 2^+$ because the differences between Lyapunov
exponents $z_{i+1}-z_i$ become very small for $\e\to 0$.

Different forms of $P(g)$ for bifractals A and B indicate that  the critical conductance distribution and also  conductance cummulants  depend on the lattice topology.
A more convincing proof of this statement is in figure 3 where we present 
the critical conductance distributions for  3D 
systems with honeycomb and triangular lattice in the $xy$ plane and compare them with $P(g)$ for 3D cubic lattice.

Figure 4 presents   $P(\log g)$ for 4D cubic lattice 
(with fixed boundary conditions). The mean conductance is half of that in 3D cubic lattice. The shape of $P(g)$ and $P(\log g)$ is similar to the one for 3D
(figure 3) and it can be analyzed by standard methods 
\cite{PRL,RMS2}.

\medskip

Finally, figure 5 summarizes the dimension dependence of the critical
parameters. 
The main difference between our numerical data  for $\nu$ and 
those published in \cite{grussbach} is the estimation of the spectral
dimension which is more accurate in the present work.
Contrary to Ref. \cite{grussbach},
we do not try to fit the dimension-dependence of our data to any 
simple function. 
As pointed out in \cite{wegnera}, $\e$-dependence of $\nu$ is non-trivial.
The leading term $\nu=1/\e$ would be observable only on lattices
with much smaller spectral dimension.
As an  exact knowledge  of $d_s$ is  crucial for  
numerical estimation of the $\e$-dependence of $\nu$,
we did not analyse statistical fractals, for which $d_s$ must be estimated by 
numerical simulations.
Another disadvantage of statistical fractals
is that due to the topology dependence of $P(g)$ we do not suppose that
system-size invariant critical conductance distribution can be found on available scales.

The dependence of the conductance cummulants on the lattice topology 
disables quantitative comparison of numerical and theoretical data.
We can only conclude that the mean conductance increases as $\e$ decreases.
To our surprise, $\langle g\rangle$ and var $g$   for 3D and 4D  seem to follow
(\ref{kumulanty}). Does it mean that the relation $n_0=1/\e$ in (\ref{kumulanty})
underestimates
the upper bound $n_0$? A positive answer could explain the  
absence of the power law decrease of $P(g)$ for large $g$.

\medskip

In conclusion, we have analyzed the critical parameters of 
the metal-insulator transition on lattices with different 
dimension and different topology. 
We confirm that the  critical exponent $\nu$ depends 
only on the spectral dimension
of lattice. This confirms  universality of the MIT. 
Our result for $\nu$ in 4D systems agrees with 
the previous numerical estimations \cite{grussbach,isa}, 
but differs considerably from theoretical expectation $\nu=1/2$ \cite{suslov}.

We present the critical conductance distribution on the lattices of 
dimension $2.226\le d_s\le 4$ and compare them with theory. 
We prove that 
two lattices with the same spectral dimension 
but different lattice topology have a different
critical conductance distribution. This prevents a quantitative comparison 
of numerical data for the conductance cummulants and for the critical disorder 
with theoretical formulae.
In agreement with 
\cite{SC}, $P(g)$ converges to the
Gaussian when dimension decreases toward the lower critical dimension $d_s=2$.
We found no power-law tail of the distribution, maybe due to the restricted 
size of the statistical ensembles.

\medskip

\noindent We thank L. \v Samaj for fruitful discussions.
This work was supported by Slovak Grant Agency VEGA, Grant n. 2/7174/20

\begin{figure}
\psfig{file=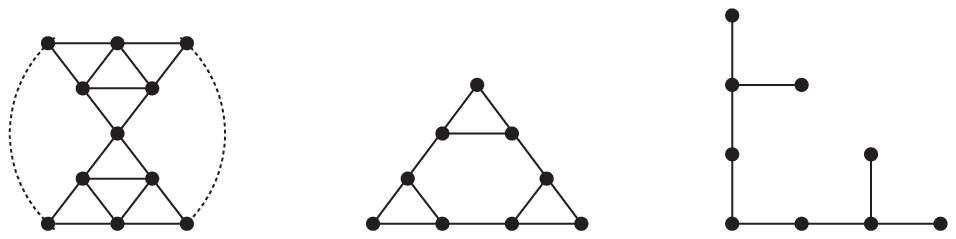,width=8cm}
\psfig{file=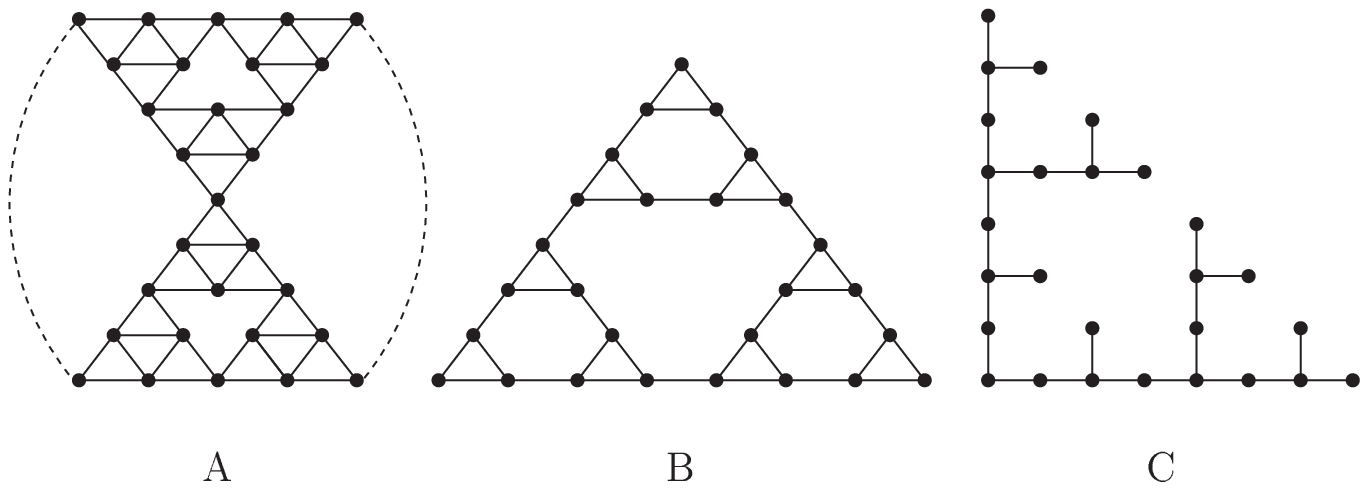,width=8cm}
\caption{The structure of three fractal lattices considered in this Letter.
Top: the second  generation (N=9), bottom: the third one (N=27).
All fractals  have the same fractal dimension
$d_f={\log 3}/{\log 2}\approx 1.58$,
A and B have  also the same spectral dimension (Table 1) \cite{exact}.
A  is doubled  Szierpinski gasket
(points connected by dotted lines are identical). B and C 
are modifications of Szierpinski gasket. Note different
number of nearest-neighbor lattice sites: it is 4 for A,  3 for B and 1, 2 or
3 for C. Note also the absence of closed loops in the C lattice.
} 
\end{figure}

\begin{figure}
%\noindent\epsfig{file=obra3.eps,width=4.3cm}~
%\epsfig{file=obrb3.eps,width=4.3cm}\\
%\epsfig{file=obrc3.eps,width=4.3cm}
%\epsfig{file=lfit.eps,width=4.3cm}
\noindent\epsfig{file=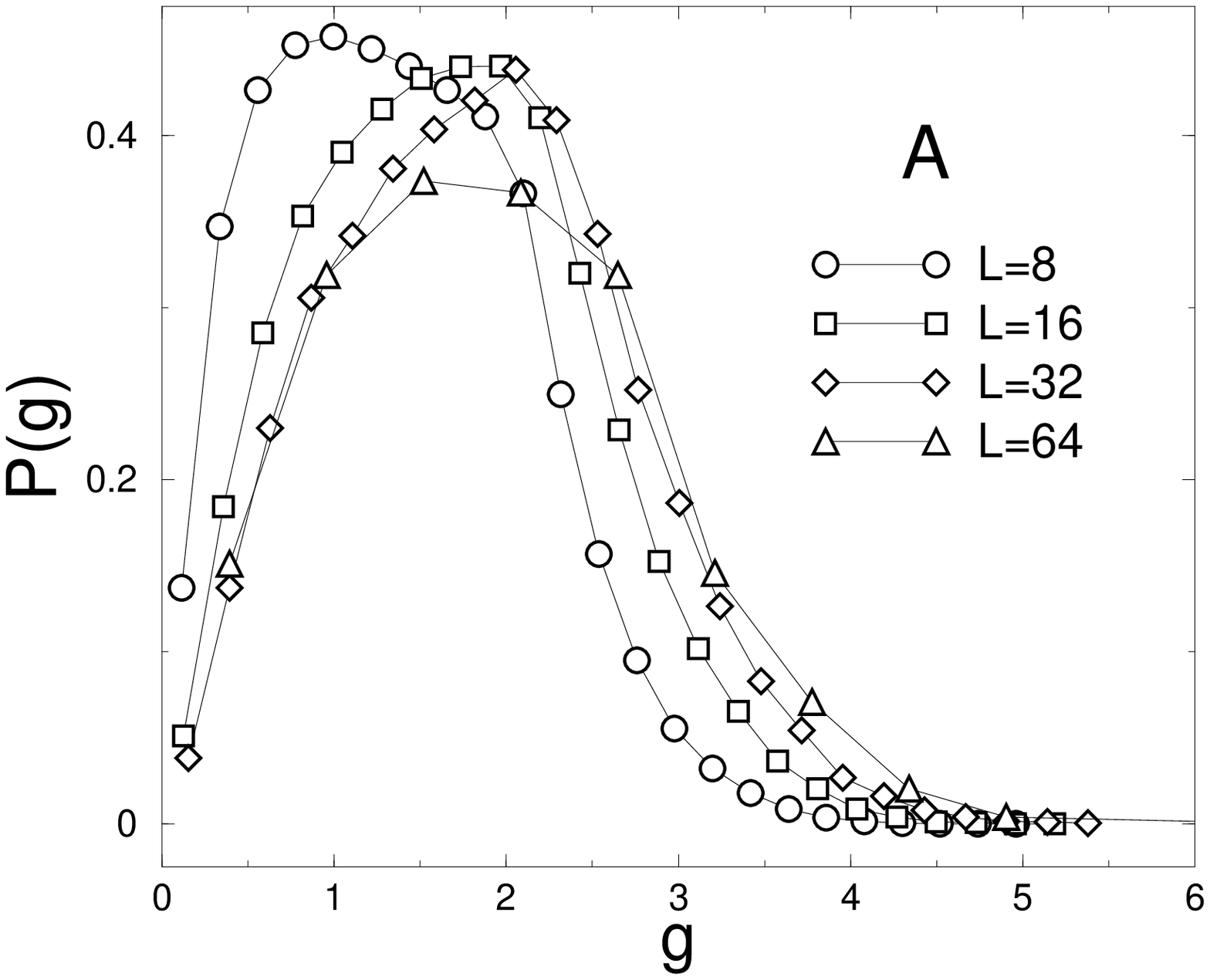,width=4.3cm}~
\epsfig{file=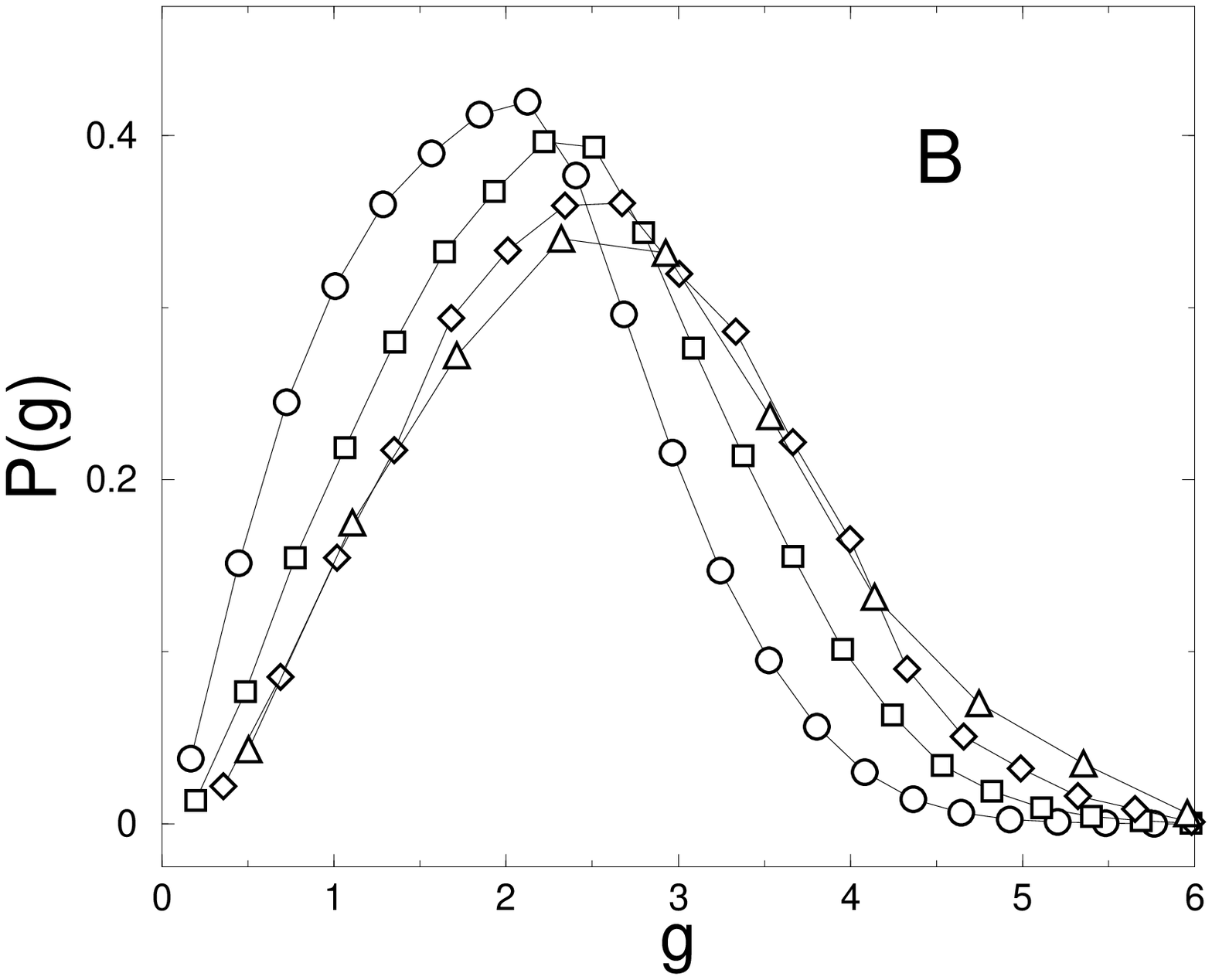,width=4.3cm}\\
\epsfig{file=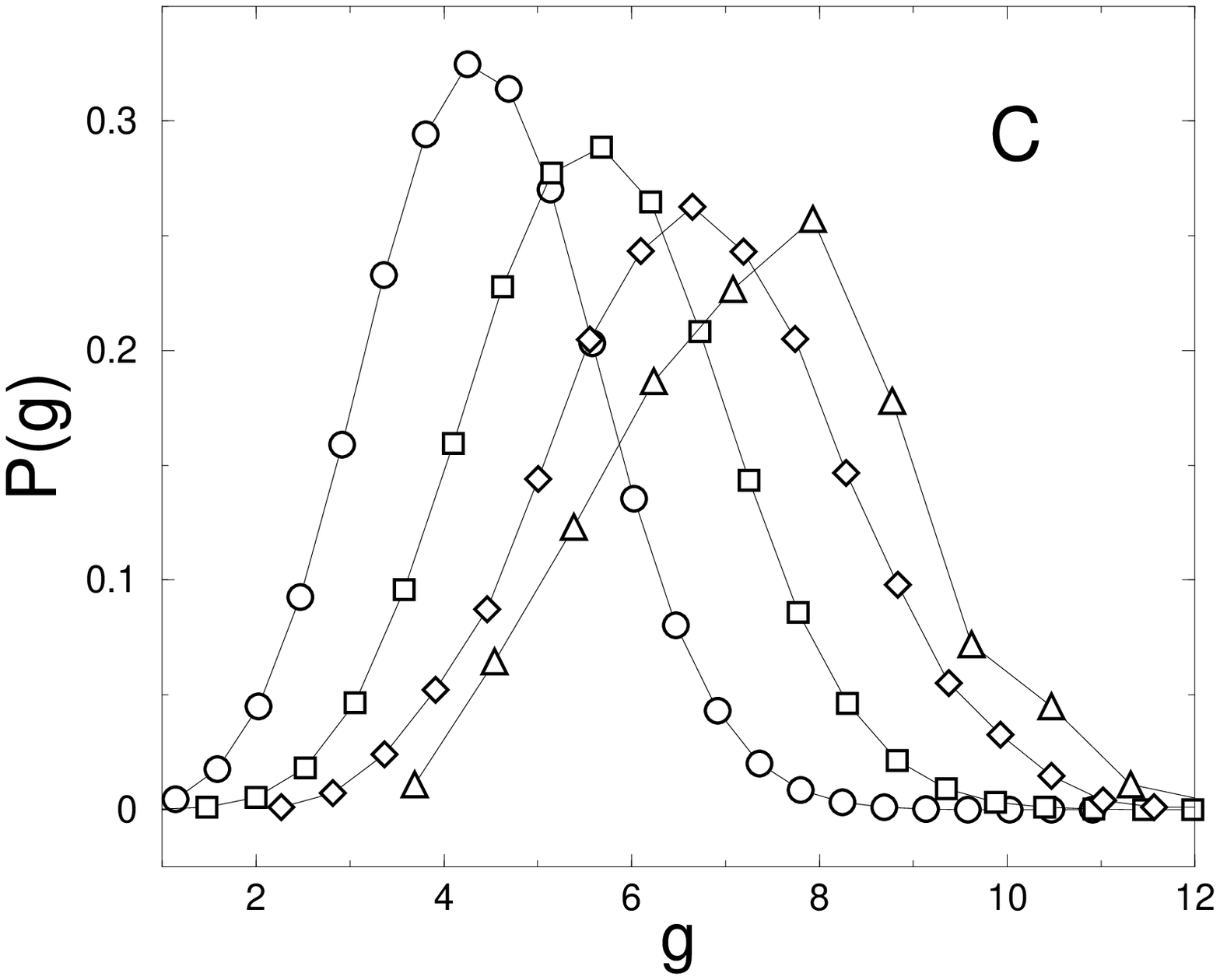,width=4.3cm}
\epsfig{file=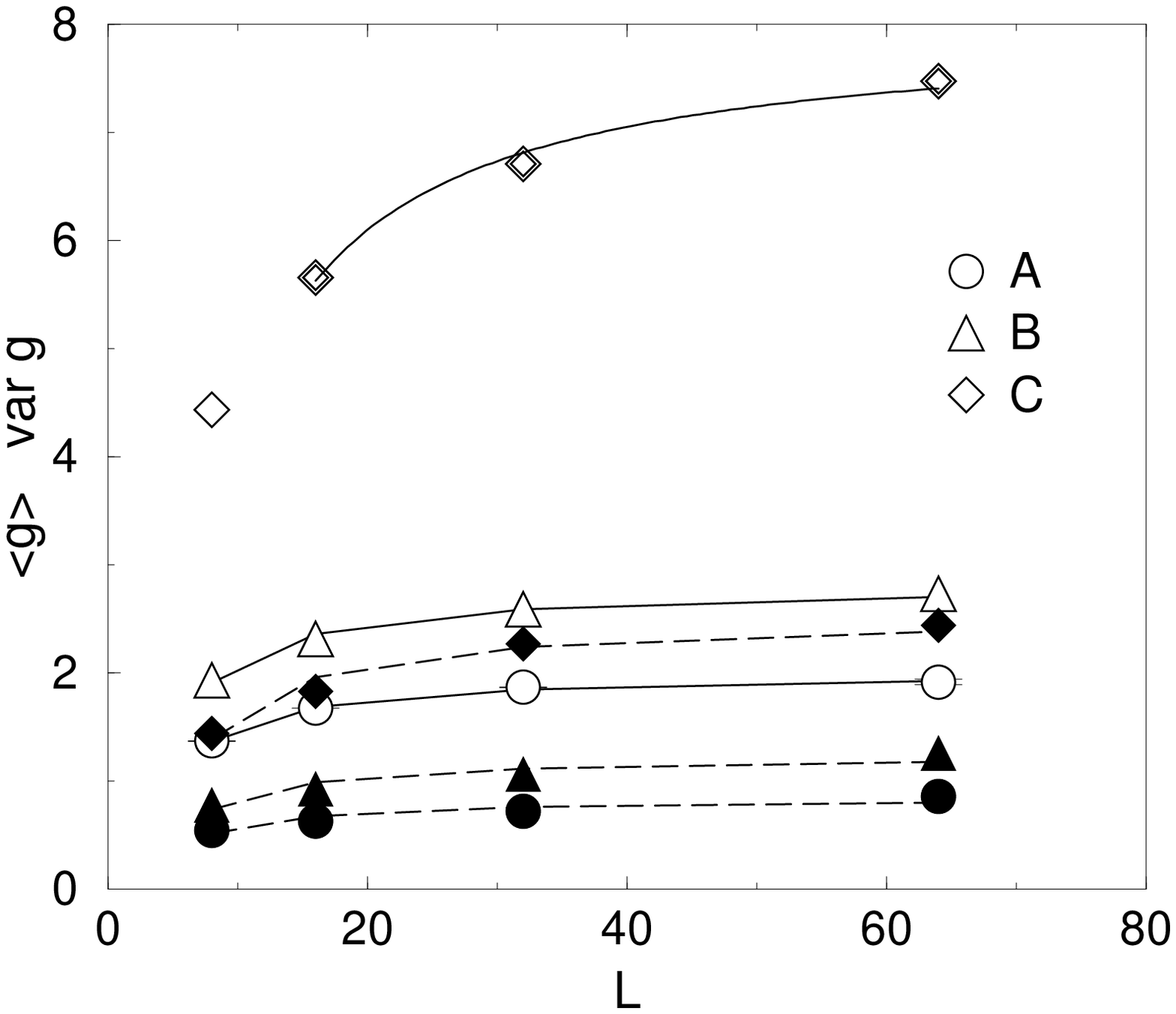,width=4.3cm}
\caption{Critical conductance distribution $P(g)$ for bifractals A, B and C. 
The number of samples in statistical ensembles  
decreases from $10^5$ for smaller systems to $\sim 1.000$ for $L=64$.
The last figure shows  the system size dependence of mean conductance (open symbols) and
var $g$ (full symbols). Note that
finite-size effects increase as critical disorder decreases. }
\end{figure}

\begin{figure}
\epsfig{file=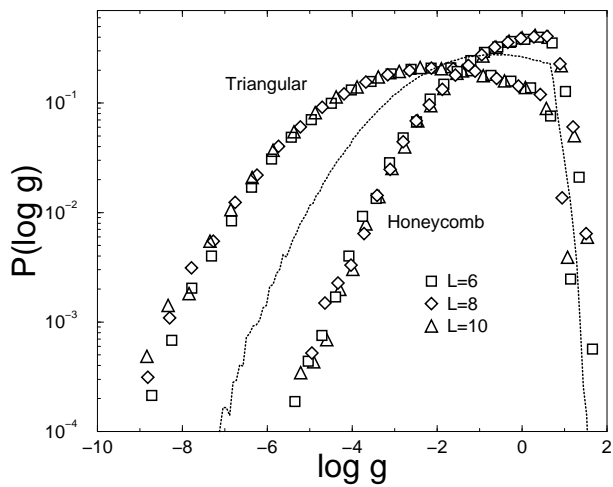,width=8cm}
\caption{Critical conductance distribution for three dimensional systems
with triangular and honeycomb structure  in the $xy$ plane.
For comparison, $P(g)$ for the cubic lattice is also presented (dotted line).}
%\epsfig{file=hexa.eps,width=8cm}\\
%\epsfig{file=triang.eps,width=6cm}
%\caption{Left panel: The first Lyapunov exponent $z_1$ as a function of disorder for different $L$ for hexagonal and triangular 3D lattice. Accuracy of data is 0.1\% for smaller $L$ and 1\% for $L=12$ and 14.  Right panel: Critical distribution for $\log g$ for the same systems. } 
\end{figure}

\begin{figure}
\epsfig{file=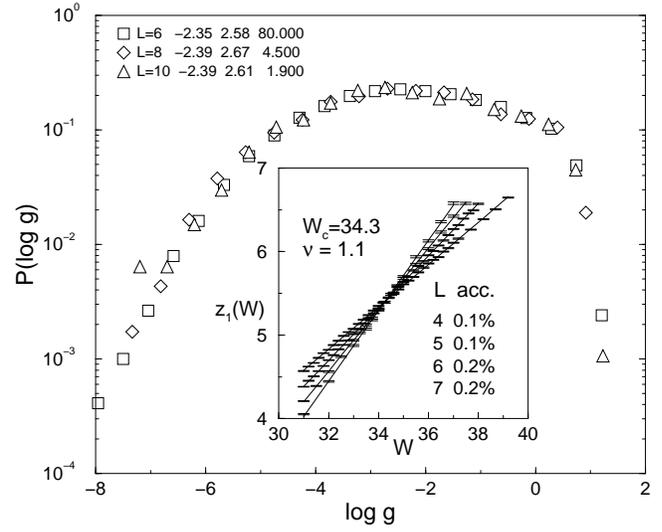,width=8.5cm}
\caption{ Critical conductance distribution of  $\log g$ for 4D hyper-cubes with fixed boundary conditions. 
The parameters of the  statistical ensembles are presented in the legend: $L$, $\langle\log g\rangle$, var log $g$ and number of samples in the statistical ensemble.
Inset: The first Lyapunov exponent $z_1$ 
as a function of disorder
for the Q1D lattice $L^3\times L_z$  
and  $4\le L\le 7$. Solid lines are linear fits $z_1(W,L)=z_1^{(0)}(L)+Wz_1^{(1)}(L)$
}
\end{figure}

\begin{figure}
\epsfig{file=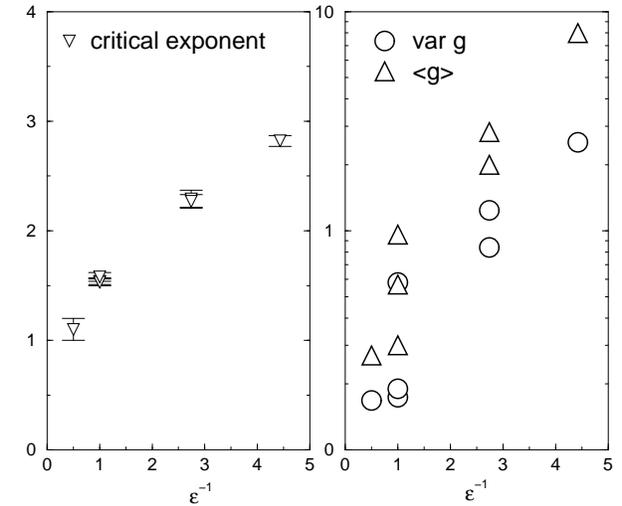,width=8cm}
\caption{Dimension dependence ($\e=d-2$) of critical exponent (left)
and of the mean and variance of conductance
(right). Systems with the same spectral dimension $d_s$ have different mean and variance of conductance, 
but they possess the same critical exponent $\nu$. For details see Tables 1 and 2.}
\end{figure}

\begin{table}
\begin{center}
\begin{tabular}{llll}
\hline
\hline
bifractal   &                    A     &     B    &    C       \\
\hline
\hline
 $d_s$    &                    2.365 &    2.365  &    2.226  \\
 $W_c$  &                        9.7(1) &   7.98(02)  & 5.77(02)  \\
  $\nu$ &                    2.27(06) & 2.29(08)   &   2.82(05)  \\ 
 $z_1$ &                         2.04(04)  &  1.59(01)  &  0.91(01)  \\ 
\hline
$\langle g\rangle$  &            2.00   &   2.82   &     7.99  \\
  var $g$  &                     0.84   &   1.24   &     2.53    \\
 $\langle\log g\rangle $  &      0.58   &   0.96   &     2.04   \\
 var log $g$  &                  0.31   &   0.22   &     0.56  \\
\hline
\hline
\end{tabular}
\end{center}
\vspace*{2mm}
\caption{Critical parameters of bifractals A, B and C (Figure 1).
$d_s$ is spectral dimension \cite{exact}. Critical disorder and critical exponent
were found from fit of the numerical data by a simple one-parameter scaling
formula.  Accuracy of critical parameters was estimated by comparison of
numerical fits for system size $L=4 \div 32$ and $L=8 \div32$ and 
for various intervals of $W$.
Limitting values of the moments of conductance  were estimated by
the fit
$X_L=X_\infty+{\rm const}/L$ with $L=8,16,32$ and 64 (figure 2).
}
\end{table} 

\begin{table}
\begin{center}
\begin{tabular}{lllll}
\hline
\hline
lattice   &                    3s    &     3h    &    3t     &  4  \\
\hline
\hline
 $d_s$    &                    3 &    3  &    3   &  4\\
 $W_c$  &                   16.5 &      13.5(2)  &  19.9(2) &   34.3(2)  \\
  $\nu$ &                    1.54(03)     &  1.58(04)  &    1.53(03)  &  1.1(1)  \\ 
 $z_1$ &                       3.45(01)   & 2.7(3)     &   4.2(2)  & 5.4(1)       \\ 
\hline
$\langle g\rangle$  &           0.57    &   0.96   &     0.30  &  0.27 \\
  var $g$  &                     0.17   &   0.58   &     0.19   &   0.17 \\
 $\langle\log g\rangle $  &      -1.93   & -0.44   &    -2.34   &  -2.39 \\
 var log $g$  &                  1.76   &   1.00   &     3.02    &  2.61\\
\hline
\hline
\end{tabular}
\end{center}
\vspace*{2mm}
\caption{Critical parameters of three dimensional systems with
square (3s) \cite{MacKK,SOK,RMS2}, honeycomb (3h) and triangular (3t) lattice in $xy$ plane
and of 4D hyper-cube. Periodic boundary conditions (b.c.) were used for Q1D simulations. 
Fixed b.c. were used  for studies 
of conductance statistics. }
\end{table}


\begin{thebibliography}{99}
\bibitem{exact} 
Spectral dimension $d_s$ is easy to find if we know how the frequency 
$\omega^2\to a\omega^2$ of lattice vibrations scales in the limit $\omega\to 0$,
 when the lattice is re-scaled as $L\to bL$. For fractal A,
$a=5$ if  $b=2$  (see Ref. \cite{spectral})  and 
$d_s^A=2\log 3/\log 5$.  We find the same $d_s$ for fractal B.
For C, 
three types of sites must be distinguished, because
the form of the  equation for lattice vibrations depends
on the number of nearest neighbors. This gives three scaling relations for
$\omega$. 
Following \cite{kva}, recurrence relations are determined by  the transformation matrix
\begin{equation}
D = \pmatrix{4 & 1 & 2 \cr 2 & 2 & 2 \cr 0 & 3 & 2 \cr}.
\end{equation}  
It larges eigenvalue is 6 and $d_s^C=2\log 3/\log 6\approx 1.226$.
\end{thebibliography}
\end{document}